\documentclass[aoas,MSNbibl,nameyear,dvips]{arximspdf}
\usepackage{dcolumn}
\usepackage[ruled,vlined]{algorithm2e}
\usepackage{graphicx}

\doi{10.1214/11-AOAS531}
\volume{6}
\issue{3}
\pubyear{2012}
\firstpage{1021}
\lastpage{1046}

\makeatletter
\newcolumntype{d}[1]{D{.}{.}{#1}}

\newcommand{\bA}{{\mathbf{A}}}
\newcommand{\bB}{{\mathbf{B}}}
\newcommand{\bC}{{\mathbf{C}}}
\newcommand{\bD}{{\mathbf{D}}}
\newcommand{\bE}{{\mathbf{E}}}
\newcommand{\bF}{{\mathbf{F}}}
\newcommand{\bG}{{\mathbf{G}}}
\newcommand{\bI}{{\mathbf{I}}}
\newcommand{\bL}{{\mathbf{L}}}
\newcommand{\bP}{{\mathbf{P}}}
\newcommand{\bQ}{{\mathbf{Q}}}
\newcommand{\bR}{{\mathbf{R}}}
\newcommand{\bS}{{\mathbf{S}}}
\newcommand{\bU}{{\mathbf{U}}}
\newcommand{\bV}{{\mathbf{V}}}
\newcommand{\bW}{{\mathbf{W}}}
\newcommand{\bX}{{\mathbf{X}}}
\newcommand{\bY}{{\mathbf{Y}}}

\newcommand{\bOmega}{\boldsymbol{\Omega}}
\newcommand{\bLambda}{\boldsymbol{\Lambda}}

\newcommand{\R}{\mathbb{R}}
\newcommand{\tr}{\operatorname{tr}}
\newcommand{\pen}{\operatorname{pen}}
\newcommand{\eqref}[1]{(\ref{#1})}
\newcommand{\fracd}[2]{({#1}/{#2})}
\makeatother

\begin{document}
\begin{frontmatter}

\title{A two-way regularization method for MEG source reconstruction}
\runtitle{Two-way regularization}

\begin{aug}
\author[A]{\fnms{Tian Siva} \snm{Tian}\corref{}\thanksref{au1}\ead[label=e1]{siva.tian@times.uh.edu}},
\author[B]{\fnms{Jianhua Z.} \snm{Huang}\thanksref{au2}\ead[label=e2]{jianhua@stat.tamu.edu}},
\author[C]{\fnms{Haipeng} \snm{Shen}\thanksref{au3}\ead[label=e3]{haipeng@email.unc.edu}}\\
\and
\author[D]{\fnms{Zhimin} \snm{Li}\ead[label=e4]{zhli@mcw.edu}}
\runauthor{Tian, Huang, Shen and Li}
\affiliation{University of Houston, Texas A\&M University, University
of North Carolina at Chapel Hill and Medical College of Wisconsin}
\address[A]{T. S. Tian\\
Department of Psychology\\
University of Houston\\
Houston, Texas 77204\\ USA \\ \printead{e1}} 
\address[B]{J. Z. Huang\\
Department of Statistics \\
Texas A\&M University \\
College Station, Texas 77843\hspace*{6pt}\\ USA \\ \printead{e2}}
\address[C]{H. Shen\\
Department of Statistics \\ \quad and Operations Research \\
University of North Carolina\\ \quad  at Chapel Hill \\
Chapel Hill, North Carolina 27599 \\ USA \\ \printead{e3}}
\address[D]{Z. Li\\ Department of Neurology \\ Medical College of
Wisconsin \\
Milwaukee, Wisconsin 53226 \\USA \\ \printead{e4}}
\end{aug}
\thankstext{au1}{Supported in part by the University of Houston New
Faculty Research Program.}
\thankstext{au2}{Supported in part by NCI (CA57030), NSF
(DMS-09-07170, DMS-10-07618) and King Abdullah University of Science
and Technology (KUS-CI-016-04).}
\thankstext{au3}{Supported in part by NIDA (1 RC1 DA029425-01) and NSF
(CMMI-0800575, DMS-11-06912).}

\received{\smonth{5} \syear{2011}}
\revised{\smonth{11} \syear{2011}}

%
\begin{abstract}
The MEG inverse problem refers to the reconstruction of the neural
activity of the brain
from magnetoencephalography (MEG) measurements. We propose a two-way
regularization (TWR)
method to solve the MEG inverse problem under the assumptions that only
a small number of locations
in space are responsible for the measured signals (focality), and each
source time course is smooth in time (smoothness). The focality and
smoothness of the reconstructed signals are ensured respectively by
imposing a sparsity-inducing penalty and a roughness penalty in the
data fitting criterion. A two-stage algorithm is developed for fast
computation, where a raw estimate of the source time course is obtained
in the first stage and then refined in the second stage by the two-way
regularization. The proposed method is shown to be effective on both
synthetic and real-world examples.
\end{abstract}

%

\end{frontmatter}
%
\section{Introduction} \label{secintro}
Magnetoencephalography (MEG) is a noninvasive neurophysiological
technique that measures the magnetic field generated by neural activity
of the brain using a collection of sensors outside the scalp [\citet
{P1995}]. When information is being processed at some regions of the
brain, small currents will flow in the neural system, producing a small
electric field, which in turn produces an orthogonally oriented small
magnetic field according to Maxwell's Equations. The MEG inverse
problem refers to recovering neural activity by means of measurements
of the magnetic field. The neural activities are usually represented by
magnetic dipoles, which are closed circulations of electric currents,
that is, loops with some constant current flowing through. Each dipole
has a position, an orientation, and a magnitude. The inverse problem
then becomes determining the position, orientation, and magnitude (or
amplitude) of the dipoles.

One challenge of the MEG inverse problem is that it does not have
a~unique solution and so it is ill-posed [\citet{vH1853};
\citet{N81};
\citet{S87}]. As
early as in the 19th century, von Helmholtz demonstrated theoretically
that general inverse problems, such as those aiming at identifying the
sources of electromagnetic fields outside a volume conductor, have an
infinite number of solutions [\citet{vH1853}]. Hence, to derive a
practically meaningful solution from the infinitely many mathematically
correct solutions, one has to introduce constraints to the solution
and/or use prior knowledge about the brain activity.

Existing approaches for the MEG inverse problem can be grouped into two
major classes that differ in how they impose constraints on the source
signals. Within the first class, the dipole fitting [\citet
{Sv86};
\citet{HHIKL};
\citet{Yamazaki00};
\citet{Jun05}] and scanning methods [\citet
{Sorrentino09};
\citet{S86};
\citet{MLL92};
\citet{VB88};
\citet{VDYS97};
\citet{DN00}] assume that there exist a
limited number of dipoles as point sources of the magnetic field in the
brain. By constraining the number of sources, the locations of these
dipoles are estimated by least squares fitting [\citet{LK03}] or
iterative computing [\citet{Baillet01}]. Dipole orientations and
amplitudes can be effectively estimated within these locations.
However, estimating the source locations involves solving a difficult
nonlinear optimization problem which has multiple local optima [\citet{DPKL04}].

Our proposed method belongs to the second class, which contains various
imaging methods. Different from the first class, imaging methods assume
that there are a large number of potential dipole locations evenly
distributed all over the cortex. By dividing the cortical region into a
fine grid and attaching a dipole at each grid, imaging methods model
the orientations and magnitudes for all the potential dipoles
simultaneously. Dipoles with nonzero magnitudes are identified as the
source dipoles. Imaging methods are based on the theory that the
primary sources can be represented as linear combinations of neuron
activities [\citet{B94}]. One can express the inverse problem using a
linear model
%
\begin{equation} \label{eqdata}
\bY=\bX\bB+\bE,
\end{equation}
where $\bY$ is an $n \times s$ matrix containing MEG time courses
measured by $n$ sensors at $s$ time points, recording the amplitudes of
the magnetic field. Without loss of generality, it is assumed that the
$s$ measurements for each time course are sampled at the same\vadjust{\goodbreak}
evenly-spaced time points. The known $n \times p$ design matrix $\bX$
links the source signals to the sensor measurements, and is computed
using a boundary element model prior to application of Model \eqref
{eqdata} [\citet{MLL99}]. The $p \times s$ matrix $\bB$ represents
the unknown dipole activities in the form of $p$ unobservable source
time courses. The $n \times s$ matrix $\bE$ contains some additive noise.
The amplitudes and orientations of the signal for each dipole at a time
point can be decomposed into three components in the $x,~y,~z$
coordinate system. Therefore, $p$ represents the total number of the
dipole components, and it is three times that of the number of grid
cells. In a typical MEG study, $s$ is usually from a few hundred to a
few thousand, $n$ is a few hundred, but $p$ is as large as over 10,000,
and so $p \gg n$.

Defining the matrix Frobenius norm as $\|\bB\|_F=\sqrt{\tr(\bB^T\bB
)}$. To recover~$\bB$, one can solve a penalized least squares problem
%
\begin{equation} \label{eqinv}
\min_{\bB} \{\|\bY-\bX\bB\|_F^2+\lambda\pen(\bB) \},
\end{equation}
where $\pen(\cdot)$ is a penalty function that promotes certain
desirable properties on $\bB$.

In the literature of MEG source reconstruction, spatial focality and
temporal smoothness are two valid assumptions. That is, the source
signals are smooth in time, and only a small number of compact areas
are responsible for the recordings [\citet{BVN09}]. Many of the imaging
methods focus on either
the first assumption or the second. Earlier methods using the
smoothness assumption usually adopt the $L_2$-norm penalty,
$\pen(\bB)=\|\bW\bB\|_F^2$ for certain weighting matrix $\bW$.
The simplest such method is the minimum norm estimate (MNE)
[\citet{HI94}] which uses $\bW={\mathbf{I}}$. The LORETA methods
[\citet{PML94};
\citet{P02}] set $\bW$ to be the discrete spatial
Laplacian operator. Two advantages of the $L_2$-penalty based
methods are the computational efficiency and the less-spiky property
in the time domain. Nevertheless, the $L_2$-penalty lowers the
spatial resolution and causes the well-known ``blurring effect''
in the spatial domain. Utilizing the $L_2$-penalty, the FOCUSS method
[\citet{GR97}] reduces the blurring effect by
reinforcing the strong signals while weakening the weak ones using
an iterative algorithm to update $\bW$. However, it is noticed that
FOCUSS is very sensitive to noise [\citet{OHG09}]. Many hierarchical
Bayesian approaches induce the temporal smoothness by employing
smoothing priors which penalize discontinuities
[see, e.g., \citet{BG97};
\citet{DMCGBL06};
\citet{NummenmaaHier07}].

An alternative penalty is the $L_1$-norm, $\pen(\bB)=|\bB|=\sum
_i^p\sum_j^s|b_{ij}|$, which promotes the focality of the recovered signals.
Related work includes the minimum current estimate (MCE) [\citet
{MO95};
\citet{UHS99};
\citet{LBDH06}] and\vadjust{\goodbreak} the sparse source imaging method [\citet
{DH08}]. In contrast to
the $L_2$-penalty, the $L_1$-penalty causes ``spiky'' discontinuities
of the recovered signals in both temporal and spatial domains.
Bayesian methods developed by \citet{BG97},
\citet{multisparse08},
\citet{sparsMEG07} take
into account the spatial focality by employing anatomic sparse priors.
However, these methods have similar problems as methods based on the
$L_1$-penalty.

To prevent the spiky property from the $L_1$-penalty and the blurry
property from the $L_2$-penalty, some $L_l$-norm methods with $0<l<1$ and
$1<l<2$ have been introduced [\citet{ANHJ05};
\citet{JLS87}]. However, the optimization
problems associated with $L_l$-penalties are more difficult to solve
than with $L_1$ and $L_2$ penalties.

More recently, some spatio-temporal regularization methods have been
proposed, which take into account both the smoothness and focality
properties by combining basis representation with penalization. The
$L_1L_2$-regulari\-zation discussed by \citet{OHG09} first projects $\bB
$ onto a temporal basis and then imposes the $L_1$-penalty on the
spatial domain and the $L_2$-penalty on the temporal domain. The event
sparse penalty procedure [\citet{BVN09}] divides the brain surface
into several patches based on its anatomic features and uses temporal
basis functions to represent source time courses within each patch. One
drawback of both methods is that it is not straightforward to choose
the basis. Both methods have some shortcomings. The former makes the
assumption that the source temporal basis
can be extracted perfectly from the MEG recordings. The latter utilizes
comprehensive prior information of the experiment task and the brain
geometry. In addition, the use of basis representation can potentially
cause information loss, since information orthogonal to the basis set can
not be recovered after the projection to the basis set is done.

The goal of this paper is to develop an innovative two-way
regularization method (TWR) for solving the MEG inverse problem
that promotes both spatial focality and temporal smoothness of the
reconstructed signals. The proposed method is a two-stage procedure.
The first stage produces a raw estimate of $\bB$ using a fast minimum
norm algorithm. The second stage refines the raw estimate in a
penalized least squares matrix decomposition framework. A
sparsity-inducing penalty and a roughness penalty are employed to
encourage spatial focality and temporal smoothness, respectively.

The proposed TWR has three major advantages over the existing methods.
First, TWR regularizes in both spatial and temporal domains, and
simultaneously takes into account both focality and smoothness
properties. Hence, it should be superior to one-way regularization
methods (e.g., MNE and MCE). Second, unlike some aforementioned
spatio-temporal methods, TWR does not rely on the choice of basis
functions. Hence, it avoids the information loss due to basis
approximation. Third, the two-stage procedure is computationally
efficient. The advantages of our method are well illustrated in the\vadjust{\goodbreak}
empirical studies, which show clearly that TWR outperforms one-way
regularization methods that focus either on the focality or the
smoothness alone, and some existing two-way spatio-temporal methods as well.

Two-way regularization techniques for matrix reconstruction
have been studied in other contexts.
\citet{HSB09} present a two-way regularized singular value decomposition
for analyzing two-way functional data that imposes separate
roughness penalties on the two domains.
\citet{WTH09} and \citet{LSHM10} develop sparse singular value
decomposition methods that impose separate sparsity-inducing
penalties on the two domains. However, to the best of our knowledge, a
two-way regularization with
the sparsity penalty on one domain and the roughness penalty on the
other domain of the data matrix has not appeared in the literature.
This paper provides a novel application of the two-way regularization
method in solving the highly ill-posed MEG inverse problem,
where different types of penalties are naturally used
to meet the dual requirements of spatial focality and temporal
smoothness on the unknown source signals.

The rest of the paper is organized as follows. Section~\ref{secmethod}
presents the details of the TWR methodology including the computational
algorithm. Through a synthetic example, Section~\ref{secsim} shows
advantages of the TWR over some existing methods for solving the MEG
inverse problem. Section~\ref{secmeg} applies the TWR to a real-world
MEG source reconstruction problem. Section~\ref{seccon} concludes the
paper with some discussion about an alternative one-step approach and
related complications.

\section{Methodology} \label{secmethod}
We propose a two-way regularization (TWR) method to regularize
the recovered signals in both spatial and temporal domains.
We adopt a penalized least squares formulation that uses
suitable penalty functions to ensure the spatial focality
and the temporal smoothness of the recovered signals.
TWR is implemented in a two-stage procedure where the first stage
produces a rough estimate of the source signals and the second stage
refines the initial rough estimate using regularization.

\subsection{Stage 1}
The goal of Stage 1 is to obtain a rough estimate of the location and
the shape of the source signals. At this stage, source information in
the data is retained as much as possible. It is natural to obtain such
a~rough estimate by solving the following minimization problem:
%
\begin{equation} \label{eqmin}
\min_{\bB}\|\bY-\bX\bB\|_F^2.
\end{equation}
Note that the forward operator $\bX$ contains the information of
positions and orientations of the dipoles, and how they are
represented at the sensor level. This information can be decomposed by
applying a singular value decomposition (SVD) to $\bX$, that is,
$\bX=\bU\bD\bV^T$, where $\bU\in\R^{n \times n}$ is an orthogonal
matrix and $\bV\in\R^{p \times n}$ is a thin (since $p \gg n$)
orthonomal matrix, such that $\bU^T\bU=\bU\bU^T=\bI$ and
$\bV^T\bV=\bI$. Then the objective function in the optimization problem
(\ref{eqmin}) becomes
\[
\|\bY-\bU\bD\bV^T\bB\|_F^2=
\|\bU^T\bY-\bD\bV^T\bB\|_F^2.
\]

Let $\tilde{\bY}=\bU^T\bY$ and $\bC=\bV^T\bB$. The minimization
problem (\ref{eqmin}) is equivalent to
%
\begin{equation} \label{eqminraw}
\min_{\bB}\|\tilde{\bY}-\bD\bC\|_F^2.
\end{equation}
Let $\tilde{{\mathbf{y}}}_i^T$ and ${\mathbf{c}}_i^T$ be the $i$th row of
$\tilde{\bY}$ and
the $i$th row of $\bC$, respectively. Since~$\bD$ is a diagonal
matrix, the minimization problem (\ref{eqminraw}) can be obtained by
separately
solving for each $i$,
\[
\min_{{\mathbf{c}}_i} \{\|\tilde{{\mathbf{y}}}_i-d_i{\mathbf{c}}_i\|^2 \},
\]
where $d_i$ is the $i$th diagonal element in $\bD$. This problem
has a unique solution $\hat{{\mathbf{c}}}_i=\tilde{{\mathbf{y}}}_i/d_i$. Then the
estimated matrix $\hat{\bC}$ with $\hat{{\mathbf{c}}}_i^T$ in the $i$th row
can be obtained. Thus, a rough estimate of $\bB$ can be obtained by solving
%
\begin{equation} \label{eqchat}
\hat{\bC}=\bV^T\bB.
\end{equation}
Note that $\hat{\bC}$ is $n\times s$, $\bV$ is $p\times n$, and $\bB$
is $p\times s$. Since $p\gg n$ and $p \gg s$,
equation (\ref{eqchat}) does not have a unique solution for
$\bB$. Any solution of \eqref{eqchat} can be written as
$\bB^\dag= \bV\hat\bC+ \bV^\perp\bF$, where $\bV^\perp$ is a
$p \times(p-n)$ orthonormal matrix whose columns are orthogonal
to the columns of $\bV$ and $\bF$ is a $(p-n)\times s$ matrix.
We pick the minimum norm solution, which is
$\hat{\bB}=\bV\hat{\bC}$.
In fact, $\hat{\bB}$ solves the following optimization problem:
\[
\min_{\bB} \|\bB\|_F^2    \qquad \mbox{subject to } \|\bY-\bX\bB\|_F=0.
\]
We can see this by noticing that $\|\bB^\dag\|_F^2 =
\|\bV\hat\bC\|_F^2 + \|\bV^\perp\bF\|^2_F \geq\|\hat\bB\|^2_F$
and the equality holds when $\bF$ is a matrix of zeros.

We call this $\hat{\bB}$ the raw estimate. Note that the raw estimate can
only recover information that lies in the column space of $\bV$, and
thus any
information orthogonal to the columns of $\bV$ is lost. Since the
columns of $\bV$ are the right singular vectors of $\bX$, the column space
of $\bV$ is equivalent to the row space of $\bX$. This
information loss can also be understood by viewing the forward
operator $\bX$ as a filter that maps the source $\bB$ to the
space of the observations, $\bY$, and so the information in the
columns of $\bB$ that is orthogonal to the rows
of $\bX$ can not be recovered. Since all imaging methods are based on
Model \eqref{eqdata}, information loss is a common problem to these
methods. This is the limitation of the MEG technology. Fortunately,
according to our experience, most important information still remains
in many real-world applications, as we will see in our real data
example. Note that the methods that require basis representation may cause
additional information loss, since any information in the columns
of $\bB$ that is orthogonal to the basis chosen will also be lost.

\subsection{Stage 2} \label{secstage2}
It is obvious that the raw estimate, $\hat{\bB}$, can be noisy. The
purpose of Stage 2 is to polish the raw estimate by incorporating the
smoothness and focality assumptions. The polished solution from this
stage is denoted as $\tilde\bB$. As we will see in the simulation
study in Section~\ref{secsim}, the shapes of the time courses in the
rows of $\hat{\bB}$ are noisy but usually follow the shapes of the
true curves, and $\hat{\bB}$ may suggest a broader range of active
regions. In a~penalized least squares framework, we apply a roughness
penalty to smooth the recovered time
courses and apply a sparsity-inducing $L_1$ penalty to refine the active
regions.

In order to apply two penalty functions to $\bB$, we first use the
two-way structure of the raw estimate and
decompose it into spatial-only and temporal-only components.
Specifically, we write $\hat{\bB}$ as
%
\begin{equation} \label{eqBapp}
\hat{\bB}=\bA\bG^T,
\end{equation}
where the matrix $\bG\in\R^{s \times q} ~(q \le s)$ contains only
the temporal features of~$\hat\bB$, and $\bA\in\R^{p \times q}$
can be
treated as the spatial coefficients.
When $q<s$, the decomposition \eqref{eqBapp} suggests a
reduced-rank representation of $\hat{\bB}$.
Our empirical studies, however, suggest that any reduced-rank representation
would lead to information loss and thus the full rank model is needed
in practice. We shall focus on the full rank model ($q=s$) for the
rest of the paper. For identification purposes, we require that $\bG$
is an
orthogonal matrix, that is, $\bG^T\bG=\bG\bG^T=\bI$.

Note that when the full rank model is used, the reconstruction error
of using $\bA\bG^T$ to represent $\hat\bB$, $\|\hat\bB- \bA\bG
^T\|_F^2$, is
exactly zero. We propose to introduce focality and smoothness
requirements on $\bA$ and $\bG$ respectively at the cost of
allowing some errors in reconstructing $\hat\bB$. In particular, we
consider the following
penalized least squares problem:
%
\begin{equation} \label{eqpen-ls}
\min_{\bA,\bG}  \{\|\hat{\bB}-\bA\bG^T\|_F^2+ \mu_1 \operatorname{pen}_1(\bA)
+\mu_2\operatorname{pen}_2(\bG) \},
\end{equation}
where $\operatorname{pen}_1(\bA)$ and $\operatorname{pen}_2(\bG)$ are appropriate penalty
functions, and $\mu_1$ and~$\mu_2$ are the corresponding penalty parameters.

To ensure the spatial focality of the recovered source signals,
we employ a~sparsity-inducing penalty on $\bA$ so that the estimated
$\bA$
is a sparse matrix, that is, a large proportion of its entries are zero.
Note that if a row of $\bA$ has all zero entries, then the
corresponding row
of $\hat\bB$ has all zero entries, indicating no signal or an inactive
location. Although other choices are possible, we use the $L_1$ penalty
$\operatorname{pen}_1(\bA) = |\bA| = \sum_{i=1}^p\sum_{j=1}^q |a_{ij}|$ to
serve our
purpose. On the other hand, to induce smoothness to the time course of
the recovered source signals, we apply a roughness penalty to the columns
of $\bG$ so that each column of $\bG$ is a smooth function of time.
Let ${\mathbf{g}}= (g_1,\ldots, g_s)^T$ represent a generic vector representing
a column of $\bG$. One choice of the roughness penalty is the squared
second order difference penalty, defined as
$\operatorname{pen}({\mathbf{g}}) = \sum_{l=2}^{s-1} (g_{l-1}- 2g_l+ g_{l+1})^2$.
This penalty
is a quadratic form and can be written as ${\mathbf{g}}^T \bOmega{\mathbf{g}}$
for a
nonnegative definite roughness penalty matrix,~$\bOmega$.
The overall penalty on $\bG$ is the summation of the penalty on each column,
$\operatorname{pen}_2(\bG) = \tr(\bG^T \bOmega\bG) = \sum_{j=1}^s {\mathbf{g}}^T_j \bOmega{\mathbf{g}}_j$.
Using the penalties defined above, the penalized least squares problem
\eqref{eqpen-ls} becomes
%
\begin{equation} \label{eqpen}
\min_{\bA,\bG} \{\|\hat{\bB}-\bA\bG^T\|_F^2+\mu_1|\bA
|+\mu_2\tr(\bG^T\bOmega\bG) \}.
\end{equation}

\subsection{Algorithm} \label{secalg}

We propose an iterative algorithm to solve (\ref{eqpen}) that
alternates the optimization with respect to $\bA$ and $\bG$.
The algorithm starts with setting the initial $\bG$ to be the
orthonormal matrix of the right singular vectors from the SVD of
$\hat{\bB}$. That is, let $\hat{\bB}={\mathbf{LTR}}^T$, where $\bL$
and $\bR$ are orthonormal matrices, and we set the initial $\bG=\bR$.

{\textit{Fix $\bG$, update $\bA$.}}
When $\bG$ is fixed as $\hat{\bG}$, the roughness penalty term
in the objective function (\ref{eqpen}) is irrelevant to the
optimization of $\bA$. Thus, updating $\bA$ reduces to solving
the problem
%
\begin{equation} \label{eqpenG}
\min_{\bA} \{\|\hat{\bB}-\bA\hat{\bG}^T\|_F^2+\mu_1|\bA
| \}.
\end{equation}
This is similar to one step of the iterative algorithm for the sparse
principal component analysis as formulated by \citet{SH08}.
Express $\bA\hat{\bG}^T$ as a summation of a serial of rank-one terms
%
\begin{equation} \label{eqone}
\bA\hat{\bG}^T=\sum^s_{j=1}{\mathbf{a}}_j\hat{{\mathbf{g}}}_j^T,
\end{equation}
where ${\mathbf{a}}_j$ and $\hat{{\mathbf{g}}}_j$ are the $j$th column of $\bA$ and
$\hat{\bG}$, respectively. Since simultaneous extracting of all the
rank-one terms is computationally expensive, we propose to obtain them
sequentially.

For the first rank-one term $(j=1)$, we solve for fixed $\hat{\mathbf{g}}_1$
%
\begin{equation} \label{eqvecG}
\min_{{\mathbf{a}}_1} \{\|\hat{\bB}-{\mathbf{a}}_1\hat{{\mathbf{g}}}^T_1\|
_F^2+\mu_1|{\mathbf{a}}_1| \}.
\end{equation}
This problem has a closed-from solution which is given below.
For the sake of notational simplicity, we drop the subscripts for now
and express the objective function of (\ref{eqvecG}) as
%
\begin{eqnarray}\label{eqseq}
&&\|\hat{\bB}-{\mathbf{a}}\hat{{\mathbf{g}}}^T\|_F^2+\mu_1|{\mathbf{a}}|\nonumber
\\[-8pt]
\\[-8pt]
&& \qquad =\sum^p_{i=1}  \Biggl\{ a_i^2\sum^s_{l=1}\hat{g}_l^2-2a_i\sum
^s_{l=1}\hat{b}_{il}\hat{g}_l+\sum^s_{l=1}b^2_{il}+\mu_1|a_i|
\Biggr\},
\nonumber
\end{eqnarray}
where $\hat{b}_{il}$ is the $(i,l)$th element in $\hat\bB$, and
$a_i$, $i=1,\ldots, p$, are the elements of the vector ${\mathbf{a}}$.
The minimization of (\ref{eqseq}) is equivalent to independently
solving~$p$ optimization problems
%
\begin{equation} \label{eqsoft}
\min_{a_i} \Biggl(a_i^2\sum^s_{l=1}g_l^2-2a_i\sum^s_{l=1}\hat
{b}_{il}g_l+\mu_1|a_i| \Biggr), \qquad i=1,\ldots,p.
\end{equation}
According to Lemma~2 of \citet{SH08}, the minimizer of each objective
function in (\ref{eqsoft}) is the soft thresholding rule
%
\begin{equation} \label{eqai}
\hat{a}_i=\operatorname{sign}(r_i)(|r_i|-\lambda)_+,
\end{equation}
where $r_i= \sum^s_{l=1}\hat{b}_{il}\hat{g}_l / \sum^s_{l=1}\hat{g}_l^2$,
and $\lambda= {\mu_1}/(2\sum^s_{l=1}\hat{g}_l^2)$. The $p$-vector
${\mathbf{a}}$ that
minimizes \eqref{eqseq} is $\hat{\mathbf{a}}= (\hat{a}_1,\ldots, \hat{a}_p)^T$.

After the first rank-one term $\hat{{\mathbf{a}}}_1\hat{{\mathbf{g}}}_1^T$ is obtained,
we find the second rank-one term by solving the following minimization
problem, while fixing $\hat{{\mathbf{g}}}_2$:
\[
\min_{{\mathbf{a}}_2} \{\|(\hat{\bB}-\hat{{\mathbf{a}}}_1\hat{{\mathbf{g}}}^T_1) - {\mathbf{a}}_2\hat{{\mathbf{g}}}_2^T\|_F^2 +\mu_1|{\mathbf{a}}_2| \}.
\]
This is the same problem as \eqref{eqvecG} except that the $\hat\bB
$ in
\eqref{eqvecG} is replaced by the residual
$\hat\bB_{\mathrm{res},1}=\hat{\bB}-\hat{{\mathbf{a}}}_1\hat{{\mathbf{g}}}^T_1$ from
the rank-one approximation. The rest
of the rank-one terms, $\hat{{\mathbf{a}}}_l\hat{{\mathbf{g}}}_l^T$, $l=3,\ldots, s$, can be
obtained sequentially in a~similar manner
by using the residuals from the lower-rank approximations.

{\textit{Fix $\bA$, update $\bG$.}}
When $\bA$ is fixed as $\hat{\bA}$, the $L_1$ penalty term
in (\ref{eqpen}) becomes constant and thus is irrelevant to
the optimization with respect to~$\bG$. The update of~$\bG$ then
solves the following problem:
%
\begin{equation} \label{eqpenA}
\min_{\bG} \{\|\hat{\bB}-\hat{\bA}\bG^T\|^2_F
+\mu_2\tr(\bG^T \bOmega\bG) \}.
\end{equation}
Since directly solving this problem is complicated, we solve for the
columns of $\bG$ sequentially. To obtain the first column of $\bG$,
we solve the problem
%
\begin{equation} \label{eqpenA1}
\min_{{\mathbf{g}}_1} \{\|\hat{\bB}-\hat{{\mathbf{a}}}_1{\mathbf{g}}^T_1\|^2_F
+\mu_2 {\mathbf{g}}^T_1 \bOmega{\mathbf{g}}_1 \},
\end{equation}
which has the solution
$\hat{\mathbf{g}}_1 = (\hat{\mathbf{a}}^T_1\hat{\mathbf{a}}_1\bI+ \mu_2 \bOmega
)^{-1} \hat\bB^T \hat{\mathbf{a}}_1$.
Let $\bOmega= \bP\bLambda\bP^T$ be the eigen-decomposition. Then
\[
\hat{\mathbf{g}}_1 = \bP(\hat{\mathbf{a}}^T_1\hat{\mathbf{a}}_1\bI+ \mu_2
\bLambda)^{-1}\bP^T \hat\bB^T \hat{\mathbf{a}}_1.
\]
To obtain an update of ${\mathbf{g}}_2$, we solve the problem
%
\begin{equation} \label{eqpenA1}
\min_{{\mathbf{g}}_2} \{\|(\hat{\bB}-\hat{{\mathbf{a}}}_1\hat{\mathbf{g}}^T_1)
- \hat{\mathbf{a}}_2{\mathbf{g}}_2^T\|^2_F + \mu_2 {\mathbf{g}}^T_2 \bOmega{\mathbf{g}}_2 \},
\end{equation}
which has the solution
%
\begin{eqnarray} \label{eqg2}
\hat{\mathbf{g}}_2 & =& (\hat{\mathbf{a}}^T_2\hat{\mathbf{a}}_2\bI+ \mu_2 \bOmega
)^{-1}(\hat{\bB}-\hat{{\mathbf{a}}}_1\hat{\mathbf{g}}^T_1)^T
\hat{\mathbf{a}}_2\nonumber
\\[-8pt]
\\[-8pt]
& =& \bP(\hat{\mathbf{a}}^T_2\hat{\mathbf{a}}_2\bI+ \mu_2 \bLambda)^{-1} \bP
^T\hat\bB_{\mathrm{res},1}^T \hat{\mathbf{a}}_2,
\nonumber
\end{eqnarray}
where again $\hat\bB_{\mathrm{res},1}=\hat{\bB}-\hat{{\mathbf{a}}}_1\hat{\mathbf{g}}^T_1$ is the
residual from the rank-one approximation. The rest of
$\hat{\mathbf{g}}_l$, $l=3,\ldots, s$, can be obtained similarly
using the residuals from the\vadjust{\goodbreak} corresponding lower rank approximations.
When all columns of $\hat\bG$ are obtained, we orthonormalize the
columns of $\hat{\bG}$ by taking the QR decomposition of $\hat\bG$
and assigning the $\bQ$ matrix to $\hat{\bG}$.


The iterative TWR procedure, including Stage 1 and Stage 2, is
summarized in Algorithm~\ref{algTWR}.

\begin{algorithm}
\dontprintsemicolon
\KwIn{$\bX,~\bY,~\mu_1,~\mu_2,~q$}
\KwOut{$\tilde{\bB}$}
\Begin{
Stage 1: \;
Obtain the SVD of $\bX$: $\bX=\bU\bD\bV^T$, $\bD=\operatorname{diag}(d_1,\ldots
,d_n)$\;
$\tilde{\bY} \leftarrow\bU^T\bY,~\tilde\bY=(\tilde
{y}_1^T,\ldots,\tilde{y}_n^T)^T$\;
\For{$i \leftarrow1$ \KwTo$n$}{
$\hat{{\mathbf{c}}}_i \leftarrow\tilde{y}_i/d_i$\;
}
$\hat{\bC} \leftarrow[\hat{{\mathbf{c}}}_1^T,\ldots,\hat{{\mathbf{c}}}_n^T]^T$\;
$\hat{\bB} \leftarrow\bV\hat{\bC}$\;
\BlankLine
Stage 2: \;
Obtain the SVD of $\hat\bB$: $\hat{\bB}={\mathbf{LTR}}^T$\;
Initialization: $\hat{\bG} \leftarrow{\mathbf{R}},~\hat\bG=(\hat
{g}_{jl})$\;
Obtain eigen-decomposition of $\bOmega$: $\bOmega=\bP\bLambda\bP
^T$\;
\Repeat{convergence of $\tilde\bB\leftarrow\hat{\bA}\hat{\bG}^T$}
{
Update $\bA$:\;
$\hat{\bB}_{\mathrm{res}} \leftarrow\hat{\bB},~\hat{\bB}_{\mathrm{res}}=(\hat
{b}_{\mathrm{res},il})$\;
$\hat{\mathbf{a}}_0\hat{\mathbf{g}}_0^T \leftarrow\mathbf{0} \in\R^{p
\times s}$\;
\For{$j \leftarrow1$ \KwTo$q$}{
$\hat{\bB}_{\mathrm{res}} \leftarrow\hat{\bB}_{\mathrm{res}}-\hat{\mathbf{a}}_{j-1}\hat
{\mathbf{g}}_{j-1}^T$\;
$\lambda_j \leftarrow\frac{\mu_1}{2\sum^s_{l=1}\hat{g}_{jl}^2}$\;
\For{$i \leftarrow1$ \KwTo$p$}{
$r_{ij} \leftarrow\frac{\sum^s_{l=1}\hat{b}_{\mathrm{res},il}\hat
{g}_{jl}}{\sum^s_{l=1}\hat{g}^2_{jl}}$\;
$\hat{a}_{ij} \leftarrow \operatorname{sign}(r_{ij})(|r_{ij}|-\lambda_j)_+$\;
}
}
$\hat\bA\leftarrow(\hat{a}_{ij})$\;
Update $\bG$:\;
$\hat\bB_{\mathrm{res}} \leftarrow\hat{\bB}$\;
$\hat{\mathbf{a}}_0\hat{\mathbf{g}}_0^T \leftarrow\mathbf{0} \in\R^{p
\times s}$\;
\For{$j \leftarrow1$ \KwTo$q$}{
$\hat{\bB}_{\mathrm{res}} \leftarrow\hat{\bB}_{\mathrm{res}}-\hat{\mathbf{a}}_{j-1}\hat
{\mathbf{g}}_{j-1}^T$\;
$\hat{{\mathbf{g}}}_j \leftarrow\bP(\hat{\mathbf{a}}^T_j\hat{\mathbf{a}}_j\bI+
\mu_2 \bLambda)^{-1}\bP^T \hat{\bB}_{\mathrm{res}}^T \hat{\mathbf{a}}_j$\;
}
$\hat\bG\leftarrow(\hat{\mathbf{g}}_1,\ldots,\hat{\mathbf{g}}_q)$\;
Obtain QR decomposition of $\hat{\bG}$: $\hat{\bG}=\bQ{\mathbf{R}}$\;
$\hat{\bG} \leftarrow\bQ$\;
}
}
\caption{The TWR algorithm}\label{algTWR}
\end{algorithm}

We consider the algorithm has converged if the Frobenius norm of the
relative difference between the current solution and the previous
solution is smaller than a prespecified threshold value. In our
implementation, we declare convergence when $\|\tilde\bB_i - \tilde
\bB_{i-1}\|_F / \|\tilde\bB_i\|_F \le10^{-6}$. Based on our
empirical studies, only a few iterations are needed to reach
convergence; 15 iterations are usually sufficient for our numerical
examples in Sections~\ref{secsim} and~\ref{secmeg}.

\subsection{Tuning parameters} \label{sectuning}

There are two tuning parameters in the TWR algorithm: the focality
parameter, $\mu_1$, and the roughness penalty parameter,~$\mu_2$. The
choice of $\mu_1$ and $\mu_2$ can be done using the cross-validation
(CV) techniques and the generalized cross-validation, respectively.

To select $\mu_1$, we can utilize the \emph{leave-one-out} CV that
minimizes the leave-one-out CV score defined as
%
\begin{equation} \label{eqcv}
\operatorname{CV}(\mu_1) = \frac{1}{n}\sum_{i=1}^n \|\bY_{i}-\bX_{i}\hat{\bA
}_{-i}\hat{\bG}^T_{-i}\|^2_F,
\end{equation}
where $\bY_{i}$ is the $i$th row of $\bY$ corresponding
to the $i$th time course, $\bX_{i}$ is the $i$th row of $\bX$,
and $\hat{\bA}_{-i}$ and $\hat{\bG}_{-i}$ are the estimates of
$\bA$ and $\bG$ using all observations except the $i$th
time course. However, practical application of the CV has
some difficulties.
The gradient-based optimization is not feasible for
minimizing the CV score since it is not a smooth function
of $\mu_1$, a~consequence of using the $L_1$ penalty.
In addition, direct computation of the CV score is costly
because of the usual large scale of the real problem.
In a typical MEG study, $n$ is over 200, $s$ is a few hundred,
and $p$ can be over 15,000.
In order to reduce the computational cost, we propose to
use the $K$-fold cross-validation. Specifically,
we divide the rows of $\bY$ and $\bX$ into $K$ about equally
sized parts and leave out one part each time for validation,
and use the rest of the parts for estimating $\bA$ and $\bG$.
The $K$-fold CV score is defined as
%
\begin{equation} \label{eqCVk}
\operatorname{CV}(\mu_1) = \frac{1}{K}
\sum_{k=1}^K \bigl\|\bY_{(k)}-\bX_{(k)}\hat{\bA}_{-(k)}\hat{\bG
}^T_{-(k)}\bigr\|^2_F,
\end{equation}
where $\bY_{(k)}$ contains the $k$th part of the rows of $\bY$,
$\bX_{(k)}$ contains the corresponding rows of $\bX$,
and $\hat{\bA}_{-(k)}$ and $\hat{\bG}_{-(k)}$ are the estimates of
$\bA$ and $\bG$ using all observations except the $k$th part of
time courses that are left out for validation.
We used $K=5$ in our implementation. To further speed up the
algorithm,
we restrict our search only in a moderate-sized set of
discrete candidate values for\vadjust{\goodbreak} $\mu_1$. Such restrictive search is
satisfactory, since we find that the results are usually not very
sensitive to mild changes of $\mu_1$ (see Sections~\ref{secsim} and
\ref{secmeg}) and thus fine tuning of $\mu_1$ is not necessary. We
used 10 different values evenly-spaced between 0 and 1 for $\mu_1$ in
our simulations and the real-world MEG example; the search range may
need to be changed for different problems.

To select $\mu_2$, note that given $\hat{\bA}$, the update of
$q$ columns of $\hat{\bG}$ can be obtained by
solving $q$ separate penalized regression problems. For the $j$th
column, the regression has $\hat\bB_{\mathrm{res},j-1}^T \hat{\mathbf{a}}_j $ as the
input, where $\hat\bB_{\mathrm{res},j-1} = \hat\bB-
\sum_{l=1}^{j-1} \hat{\mathbf{a}}_l\hat{\mathbf{g}}_l^T$,\vspace*{2pt}
$\hat{\mathbf{g}}_j$ as the output, and the hat matrix of the regression
is $\bS_j=\bP(\hat{\mathbf{a}}^T_j\hat{\mathbf{a}}_j\bI+ \mu_2 \bLambda
)^{-1}\bP^T$,
according to equation (\ref{eqg2}). Theoretically, $\mu_2$ can take
different values for different $\hat{\mathbf{g}}_j$'s, but we
decide to use a common $\mu_2$ for all the $\hat{\mathbf{g}}_j$'s based on
computational efficiency consideration. The advantage of this
strategy is that there is only one optimization
problem to solve for choosing the tuning parameter when updating~$\hat\bG$. Then, the overall GCV criterion
is the average of all individual GCV criteria:
%
\begin{equation} \label{eqgcv}
\operatorname{GCV}(\mu_2) = \frac{1}{s}\sum^s_{j=1}\frac{\|\hat\bB
_{\mathrm{res},j-1}^T\hat{\mathbf{a}}_{j-1}-\hat{\mathbf{g}}_j\|^2}{\{1-\fracd{1}{s}\tr
(\bS_j)\}^2},
\end{equation}
where $\hat\bB_{\mathrm{res},0}=\hat\bB$, and $\tr(\bS_j) = \sum^p_{l=1}
1/\{\hat{a}_{lj}^2+\mu_{2}\lambda_l\}$.
The GCV optimization is nested in the iterations
because it is defined conditioning on the current value of~$\hat\bA$. Since the GCV
criterion is a smooth function of $\mu_2$, the optimization
can be done using a combination of golden section search and
successive parabolic interpolation [\citet{Brent73}].

\subsection{One-way regularization} \label{secowr}
By separately setting one of the penalty parameters in \eqref{eqpen}
to be zero,
one can obtain two different one-way regularization methods: tOWR and
sOWR, as explained below. These two one-way regularization methods will
be used as a comparison to TWR to demonstrate the need for two-way
regularization.

Letting $\mu_1=0$ leads to a method that emphasizes temporal
smoothness of the recovered signals, which is referred to
as tOWR (temporal one-way regularization), and is related to the
functional PCA
[\citet{HSB08}]. The corresponding optimization problem becomes
%
\begin{equation} \label{eqTWRsmooth}
\min_{\bA,\bG} \{\|\hat{\bB}-\bA\bG^T\|_F^2+\mu_2\tr(\bG
\Omega\bG^T) \}.
\end{equation}
A modified version of Algorithm~\ref{algTWR} can be applied
for computation, with the
``Update $\bA$'' step in the algorithm simplified to
$\hat{\bA}=\hat{\bB}^T\hat{\bG}$.

Letting $\mu_2=0$ leads to a method that encourages spatial
sparsity of the recovered signals, which is referred to as
sOWR (spatial one-way regularization) and is related to the sparse principal
component analysis of \citet{SH08}.
In this case, the optimization problem (\ref{eqpen}) reduces to
%
\begin{equation} \label{eqTWRsparse}
\min_{\bA,\bG} \{\|\hat{\bB}-\bA\bG^T\|_F^2+\mu_1|\bA
| \}.\vadjust{\goodbreak}
\end{equation}
Again, a modified version of Algorithm~\ref{algTWR} is applicable,
but with the ``Update~$\bG$'' step simplified to $\hat\bG= \hat{\bB
}^T\hat{\bA}$.


\section{Synthetic example} \label{secsim}
In this section we illustrate the proposed TWR method using a synthetic
example that mimics human brain activities. Both the source and the
forward operator are created based on real-world MEG studies.

\subsection{Data generation}
\begin{figure}

\includegraphics{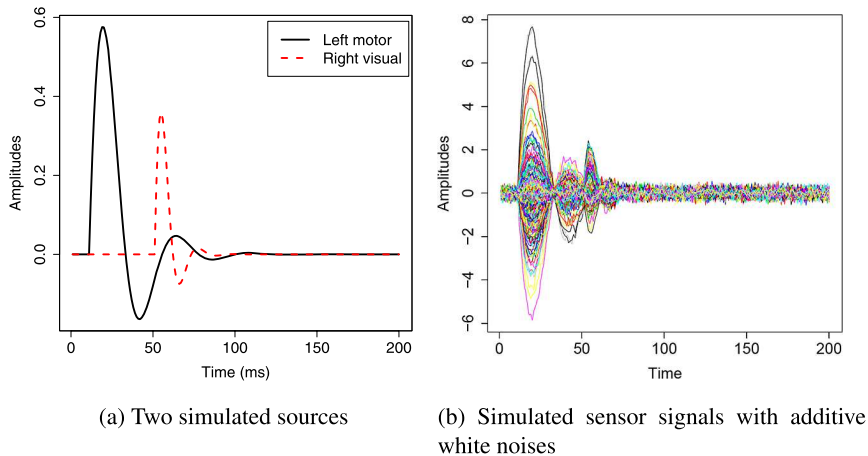}

\caption{Simulated source and sensor data.}
\label{figBsim}
\end{figure}

$\!\!\!\!$We generated the forward operator, $\bX$, from a~\mbox{human} subject
head boundary element model using the MNE software (available at:
\url{http://www.nmr.mgh.harvard.edu/martinos/userInfo/data/sofMNE.php}).
The $\bX$ matrix is a $248 \times15\mbox{,}360$ matrix,
corresponding to a MEG device with 248 valid channels.
To mimic real-world scenarios and ensure enough difficulty of
the problem, we located two source areas on the left and the right
hemispheres, respectively. The sources were
generated from two sine-exponential functions [\citet{BVN09}] and are
shown in Figure~\ref{figBsim}(a). The black solid and the red dashed
curves are source signals located at the left motor and the right
visual cortical areas, respectively. As we can see, the sources reach
their energy peaks at 25 ms and
58~ms, respectively. The synthetic MEG time courses were generated using
equation~\eqref{eqdata} and were obtained using a sampling frequency
355 Hz with a duration of 200 seconds [see Figure~\ref{figBsim}(b)].
By mimicking the real MEG data after preprocessing, that is, denoising
and smoothing, the signal-to-noise ratio,
$\mathrm{SNR} = \|\bX\bB\|^2_F/\|\bE\|^2_F $, is set to be 5dB.

%

\subsection{Comparison criteria}
We compare TWR with eight different methods that can be put into two
categories as given below.
\begin{itemize}
\item One-way regularization:
\begin{itemize}
\item The $L_2$-based MNE method [\citet{MLL99}]
\item The $L_1$-based MCE method [\citet{MO95}]
\item sOWR (i.e., spatial sparsity only)
\item tOWR (i.e., temporal smoothness only)
\end{itemize}
\item Two-way regularization:
\begin{itemize}
\item The $L_1L_2$ method proposed by \citet{OHG09}
\item MNE$+$sOWR (i.e., obtaining the MNE solution as Stage 1 and then
applying sOWR)
\item MCE$+$tOWR (i.e., obtaining the MCE solution as Stage 1 and then
applying tOWR)
\item MNE$+$TWR (i.e., obtaining the MNE solution as Stage 1 and then
applying Stage 2 of TWR)
\end{itemize}
\end{itemize}

We put MNE+sOWR in the two-way regularization category because the~$L_2$
penalty in MNE puts constraints on both domains, and sOWR puts
the~$L_1$ penalty only on the spatial domain. As a result, the temporal
domain is regularized by the $L_2$ penalty, while the spatial domain is
regularized first by the $L_2$ penalty and then by the $L_1$ penalty.
Similarly, MCE+tOWR is also categorized as a two-way regularization
method. MNE+sOWR and MCE+tOWR can be considered as two alternative ways
for two-way regularization and are suggested by a reviewer. MNE+TWR,
also suggested by a reviewer, is a slight modification of TWR,
replacing the first stage of TWR by MNE. Its inclusion in comparison
helps us study the effect of using a different Stage 1 estimator on the
performance of TWR. We implemented all the methods in R, and the tuning
parameters are selected using either CV or GCV.

Three comparison criteria are utilized: the overall mean squared error
(MSE), the standardized distance between the energy peak of the
estimated source and the energy peak of the true source, and the
computation time.

The overall MSE is defined as
\begin{displaymath}
\mathit{MSE} = \frac{1}{p}\|\bB-\tilde{\bB}\|^2_F,
\end{displaymath}
where $\bB$ and $\tilde\bB$ are the true and recovered source matrices,
respectively.

%
\begin{table}
\tabcolsep=0pt
\caption{Comparison of nine methods using four criteria: the mean
squared error (MSE), the~standardized distance between the true
energy peak and the estimated energy peak~at~the left motor area
($d_{25}$), at the right visual area ($d_{58}$), and the computation
time~(in seconds). Reported are the average and standard error
of each criterion based~on~100 simulation runs}
\label{tblcompare}
\begin{tabular*}{\textwidth}{@{\extracolsep{\fill}}ld{3.7}d{3.6}d{3.7}d{9.4}@{}}
\hline
\textbf{Method} & \multicolumn{1}{c}{\textbf{MSE ($\boldsymbol{10^{-3}}$)}}
 & \multicolumn{1}{c}{$\boldsymbol{d_{25}}$ \textbf{($\boldsymbol{\times10^{-4}}$)}}
 & \multicolumn{1}{c}{\textbf{$\boldsymbol{d_{58}}$
($\boldsymbol{\times10^{-4}}$)}} & \multicolumn{1}{c@{}}{\textbf{Computation time (sec.)}}\\
\hline
MNE & 544.0\ (9.0) & 50.2\ (7.3) & 42.9\ (5.9) & 4371\ (4.3) \\
MCE & 903.7\ (8.9) & 337.1\ (6.4) & 156.1\ (11.4) & 1545\ (3.0) \\
tOWR & 407.9\ (8.9) & 40.2\ (5.8) & 39.6\ (4.3) & 1841\ (3.4)\tabnoteref{tb1} \\
sOWR & 153.2\ (7.7) & 19.3\ (4.6) & 13.9\ (3.9) & 1798\ (3.6)\tabnoteref{tb1} \\
TWR & \multicolumn{1}{c}{\textbf{22.3\ (5.7)}\hphantom{0}} & \multicolumn{1}{c}{\textbf{15.7\ (3.3)}} & \multicolumn{1}{c}{\textbf{7.1\ (2.4)}}& 1872\ (3.5)\tabnoteref{tb1} \\
$L_1L_2$& 44.3\ (7.1) & 31.0\ (6.1) & 17.8\ (2.3) & 40\mbox{,}872\ (8.8) \\
MNE+sOWR& 187.3\ (8.8) & 27.9\ (6.8) & 14.5\ (3.1) & 5998\ (3.9) \\
MCE+tOWR& 912.7\ (10.9) & 343.8\ (6.2)& 145.2\ (12.7)& 3321\ (3.8) \\
MNE+TWR &28.6\ (7.2)& 16.9\ (4.3)&10.7\ (3.9) & 6201\ (3.1) \\
\hline
\end{tabular*}
\tabnotetext{tb1}{The computation time for each simulation run is computed
based on 15 iterations, which are usually more than needed
for algorithm convergence.}
\end{table}

The energy of the dipole $j$ at time point $k$ is defined as
$(b_{jk,x}^2+b_{jk,y}^2+b_{jk,z}^2)^{1/2}$, where
$b_{jk,x},~b_{jk,y},~b_{jk,z}~(j=1,\ldots,p,~k=1,\ldots,s)$, are the
amplitude components for the $j$th dipole at the time point $k$ in the
Cartesian coordinate system. The energy of the reconstructed source can
be defined similarly. The standardized distance between the estimated
and the true energy peak at time point $k$ is defined as
\[
d_k=\frac{\sqrt{(x^*_k- \hat{x}_k)^2+(y^*_k-\hat
{y}_k)^2+(z^*_k-\hat{z}_k)^2}}{p/3},
\]
where $p/3$ is the total number of dipoles,
$x^*_k,~y^*_k,~z^*_k$ are the coordinates of the location
for the maximum source energy at time point $k$, and $\hat{x}, ~\hat
{y},~\hat{z}$ are the coordinates for the maximum
estimated source energy at the corresponding time point. In this
simulation example, there are two peak times, 25 ms and 58 ms, so we
are interested in $d_{25}$ and $d_{58}$.


\subsection{Results}
The simulation was conducted 100 times with the noise term in
Model \eqref{eqdata} newly generated for each run. The criteria
described in the previous subsection (i.e.,
MSE, $d_{25}$, $d_{58}$, computation time) were evaluated for each
simulation run, and the mean and standard error of the criterion values
across the 100 runs were calculated.
The numerical results are shown in Table~\ref{tblcompare}.

Several interesting observations can be made from the table. TWR is the
best method in the sense of having the smallest MSE and the shortest
distances between the true and the estimated peaks. Among the four
one-way regularization methods, sOWR and tOWR outperform the classical
MNE and MCE methods, and tOWR outperforms sOWR. The fact that TWR
outperforms the four one-way regularization methods justifies our
proposal of using two-way regularization. The $L_1L_2$ method is the
third most accurate method, but its computation time is more than 21
times as large as that of TWR. MNE+sOWR and MCE+tOWR are less
satisfactory, demonstrating the importance of the first stage. MNE+sOWR
is not better than sOWR because the $L_2$ penalty of MNE does not
smooth the temporal domain. The performance of MCE+tOWR is similar to
MCE and is not better than tOWR\vadjust{\goodbreak} because MCE does not recover well
important information at the first stage, and hence tOWR based on MCE
is inaccurate. Note that the reported computation time for TWR, sOWR
and tOWR are based on fixed 15 iterations in order to make the
calculation of the average computation time meaningful. Such report is
conservative because these algorithms usually converge rapidly and
fewer iterations (usually less than 10) are enough to obtain
considerably good accuracy.

\begin{figure}

\includegraphics{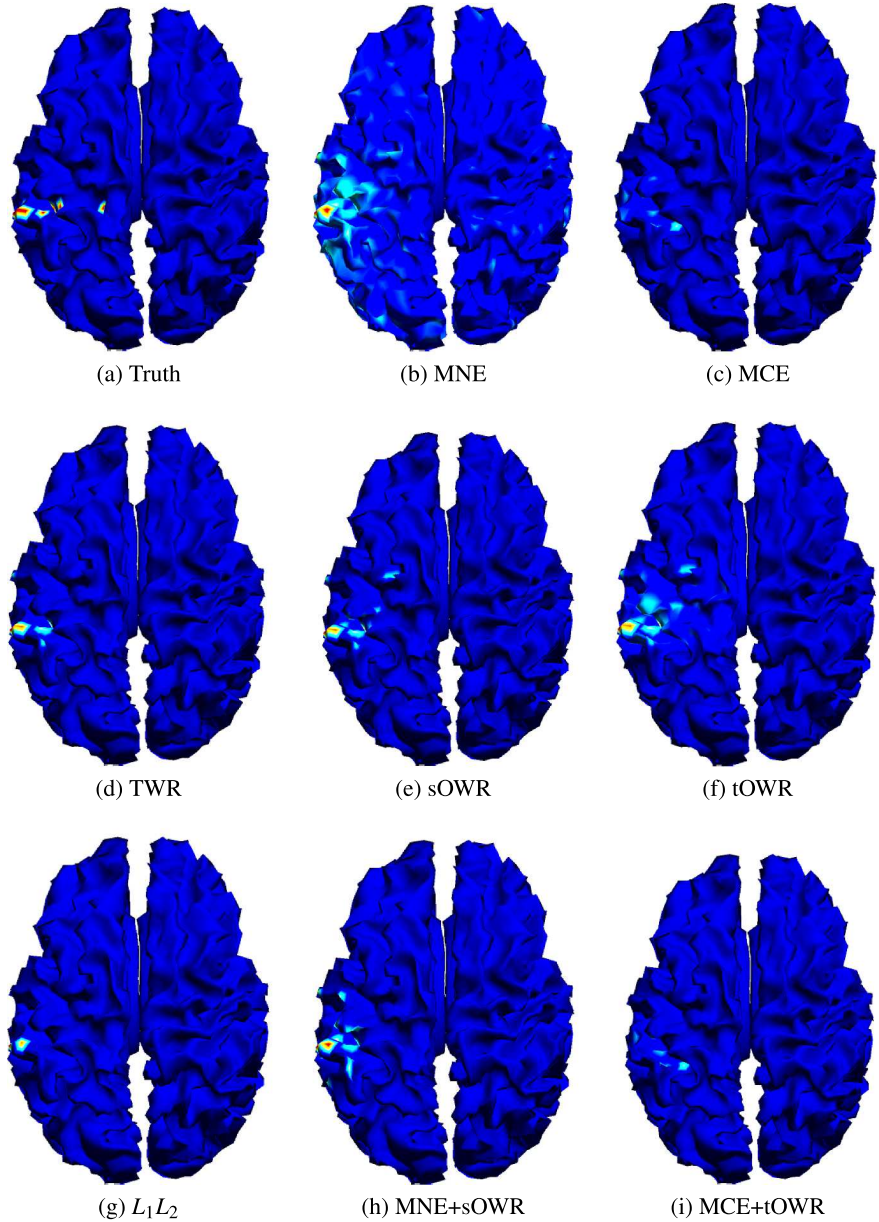}

\caption{Overviews of brain mapping by different methods at 25
ms. \textup{(a)} shows the true map, indicating an active area located at
the left motor area. TWR identifies the major active area and the
solution is focal. The $L_1L_2$ method also identifies
the active area but the solution is too focal. sOWR and MNE+sOWR
produce more scattering solutions than
TWR. MNE and tOWR detected active areas are diffuse. MCE and MCE+tOWR
misidentify the active region.}
\label{fig25}
\end{figure}

\begin{figure}

\includegraphics{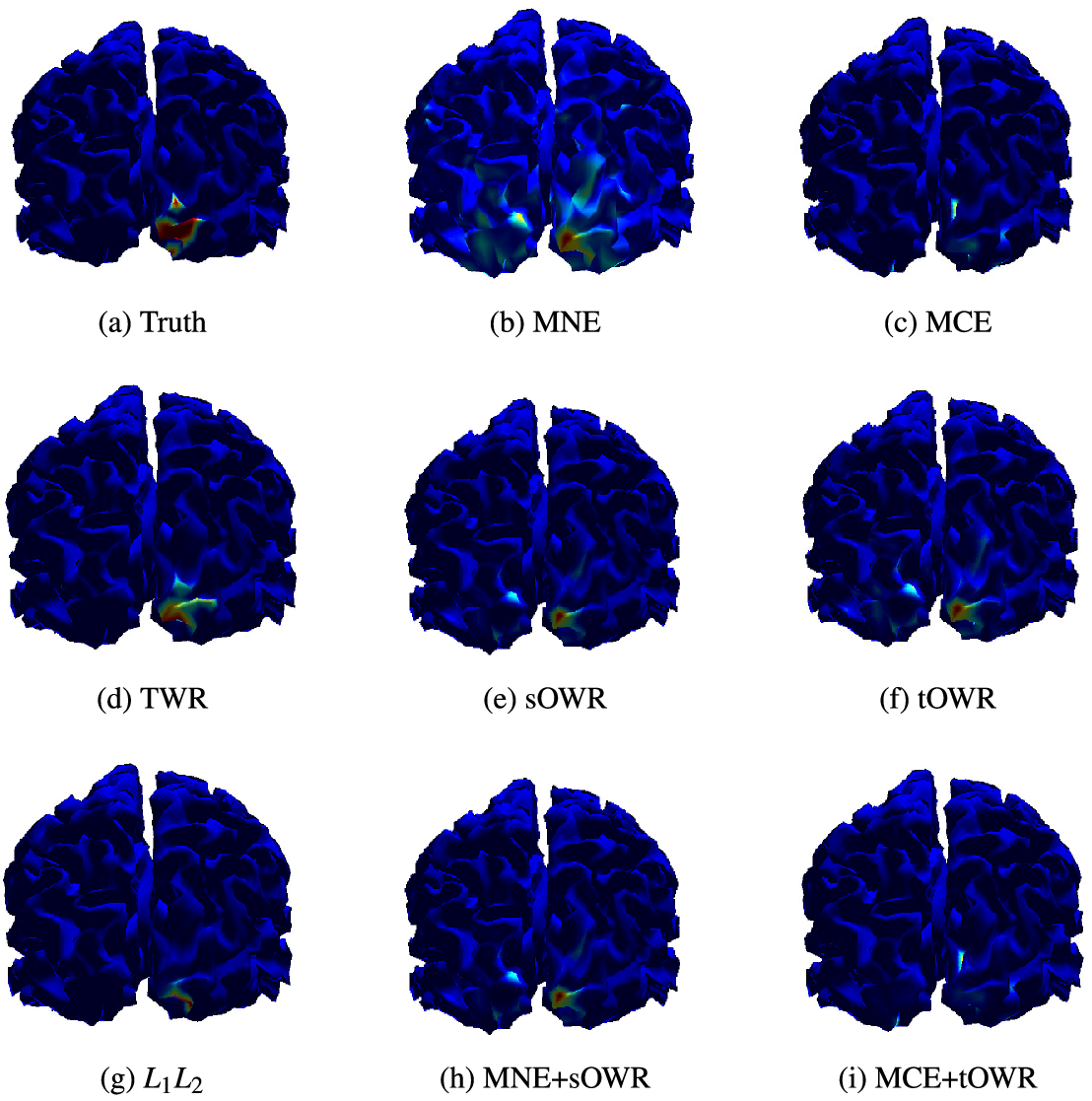}

\caption{Sideviews of brain mapping by different methods at 58
ms. \textup{(a)} shows the true map, indicating an active area located at
the right visual area. TWR and $L_1L_2$ identify the major active area
and the
solution is focal. sOWR and MNE+sOWR produce more scattering solutions than
TWR. MNE and tOWR detected active areas are diffuse. MCE and MCE+tOWR
misidentify the active region.}
\label{fig58}\vspace*{-3pt}
\end{figure}

Figures~\ref{fig25} and~\ref{fig58} show the 3-D brain
mapping by different methods at 25 ms and 58 ms for a randomly selected
simulation run.
TWR performs the best among the nine methods in detecting the true source
locations even though it misses some small regions. It is able to
identify the majority parts of both
source locations, and its solutions are focal. Solutions from sOWR and
MNE+sOWR are more scattered
than TWR. MNE and tOWR produce even more diffuse solutions.
MCE misses the main parts of both active areas and so does MCE+tOWR,
and they are the least
satisfactory methods. The $L_1L_2$ method recovers some of the
activity, but the solution is overly focal. The plot of MNE+TWR is very
similar to that of TWR, so it is not presented here to save space.
Direct comparison of results of TWR and tOWR clearly demonstrate the
positive effect of using regularization in the spatial domain.

%

%

\begin{figure}

\includegraphics{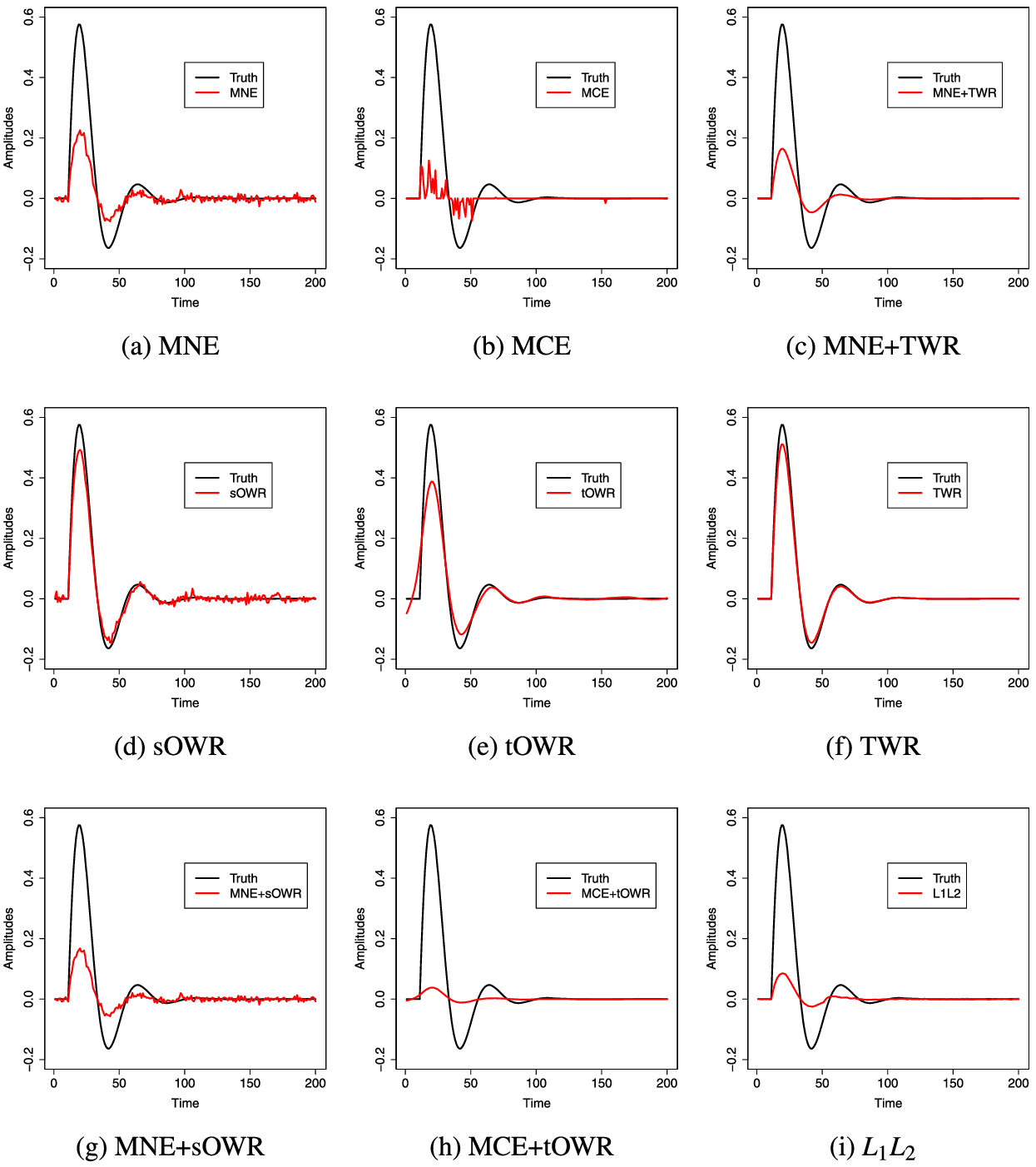}

\caption{Estimated time courses for one arbitrarily chosen dipole
component at left motor area by different methods for a randomly
selected simulation run. TWR, tOWR, MCE+tOWR, $L_1L_2$ and MNE+TWR
recover the shape of the time course reasonably well and the solutions
are smooth. But MCR+tOWR, MNE+TWR and $L_1L_2$ overshrink the
amplitude. MNE, MCE, sOWR and MNE+sOWR estimate the general trend
reasonably well, but the estimated time courses are too noisy. TWR
gives the best result.}
\label{figsource1}\vspace*{3pt}
\end{figure}

\begin{figure}

\includegraphics{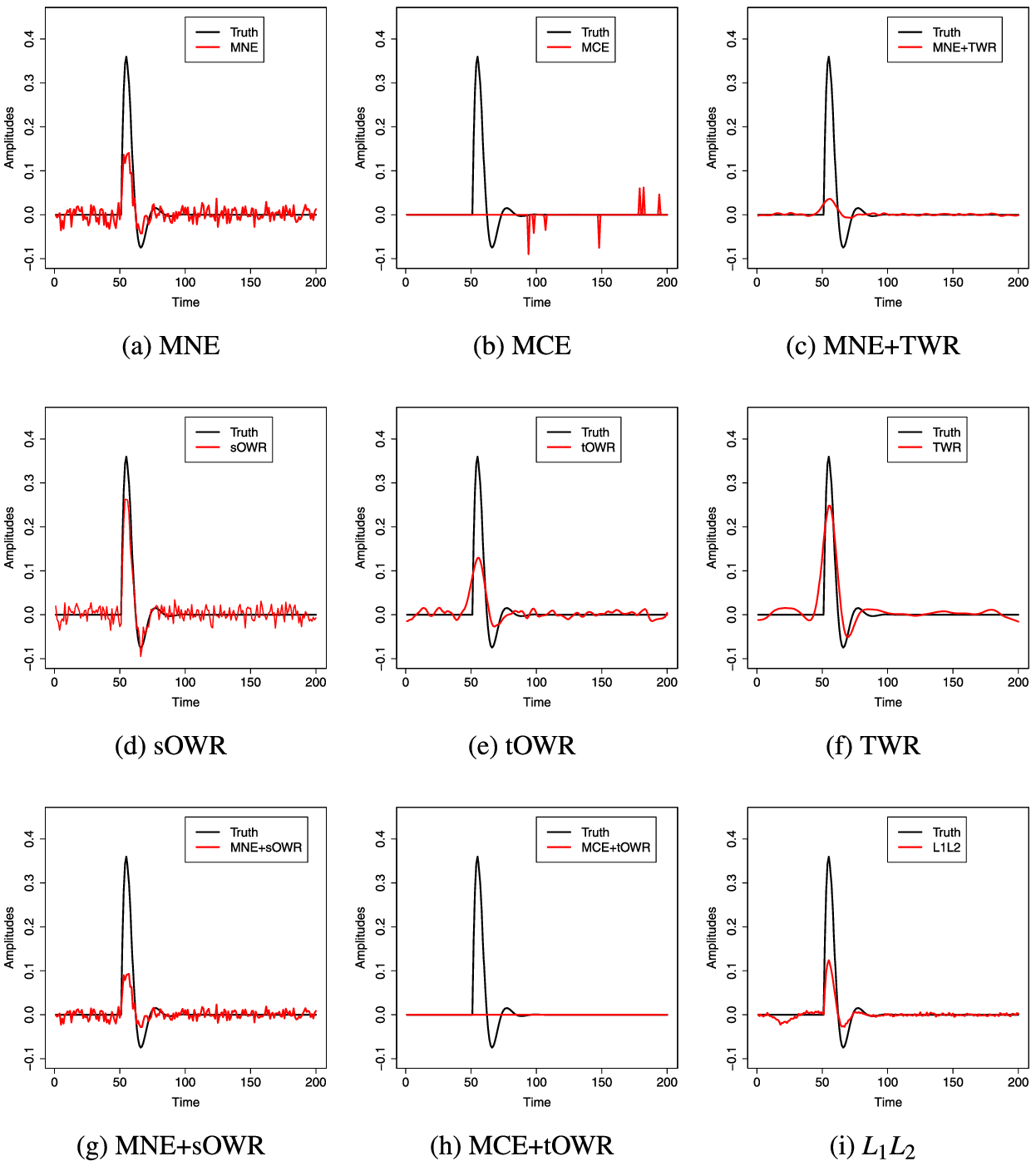}%
\caption{Estimated time courses for one arbitrarily chosen dipole
component at right visual area by different methods for a randomly
selected simulation run. TWR, tOWR, $L_1L_2$~and MNE+TWR recover the
shape of the time course reasonably well and the solutions are smooth.
But MNE+TWR overshrinks the amplitude. MNE, sOWR and MNE+sOWR estimate
the general trend reasonably well, but the estimated time courses are
too noisy. MCE and MCE+tOWR do not recover the shape of the time
course. TWR gives the best result.}
\label{figsource2}
\end{figure}

Figures~\ref{figsource1} and~\ref{figsource2} show the true and the
recovered time courses by the nine methods for an arbitrarily chosen
single dipole component in the two active areas, respectively, for a
randomly selected simulation run. Each subfigure shows the true time
course and the estimated time course by one method.
As one can see, the methods considering the temporal smoothness
reconstruct the shape of the source time course well. TWR, tOWR,
$L_1L_2$, MCE+tOWR and MNE+TWR all produce smooth time courses. TWR
recovers the most energy of the source, while MCE+tOWR recovers the
least. MNE+TWR tends to overshrink the amplitude of the time course
because MNE overshrinks the amplitude. The methods without considering
the roughness regularization in the temporal domain result in noisy
time courses even though some methods can recover the general trend. In
Figure~\ref{figsource2}(b), MCE does not capture the major peaks of
the signal, and, consequently, MCE+tOWR [Figure~\ref{figsource2}(h)], which relies on the solution of MCE, does not recover any signal
activity either. Direct comparison of results of TWR and sOWR clearly
demonstrate the positive effect of using regularization in the time domain.

%

%

\begin{figure}

\includegraphics{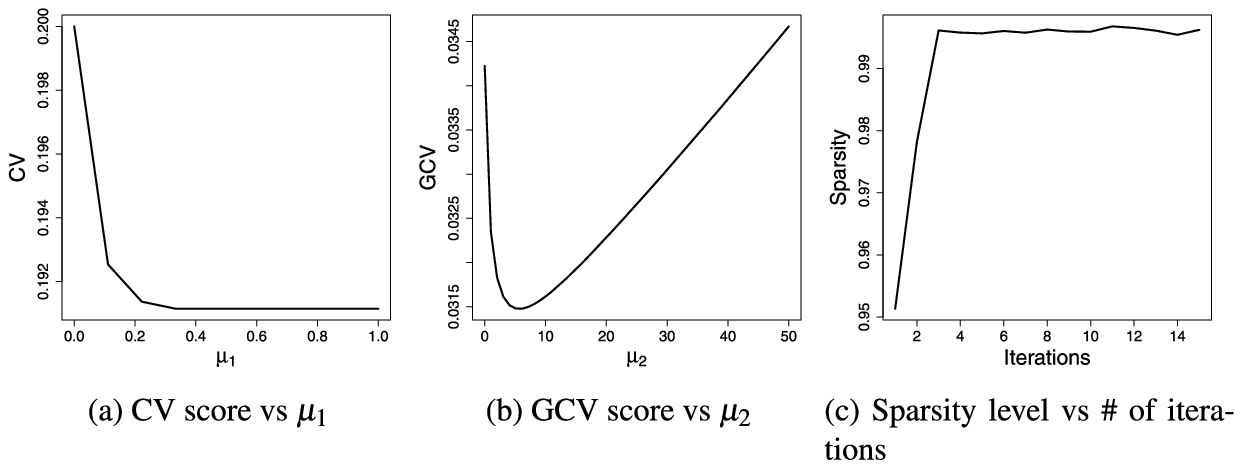}

\caption{Selection of $\mu_1$ and $\mu_2$ and the sparsity level
as a function of the number of iterations. The
optimal $\mu_1$ and $\mu_2$ are around 0.33 and 5.9,
respectively. The sparsity measure levels off at around 0.996.}
\label{figtune}\vspace*{-3pt}
\end{figure}

The selection of the focality parameter and the roughness
penalty parameter was conducted using the method presented in Section
\ref{sectuning}. Figure~\ref{figtune}(a) and (b) shows the CV
and GCV scores for TWR as functions of $\mu_1$ and $\mu_2$,
respectively. The
optimal values of the tuning parameters are $\mu_1=0.33$ and $\mu
_2=5.9$. Figure~\ref{figtune}(c) shows the sparsity level of the reconstructed
source matrix, $\tilde{\bB}$, for TWR as a function of the number of
iterations when the tuning parameters are set at the selected values.
The sparsity level for a~matrix is defined as the number of zero
entries over the total number of entries. Here the total number of\vadjust{\goodbreak}
entries for $\tilde{\bB}$ is $p\times s=3\mbox{,}072\mbox{,}000$.
From this figure, we observe that the sparsity of $\tilde{\bB}$ levels
off rather rapidly and stays steadily at about 0.996, a fairly high
sparsity level. In fact, this sparsity level matches closely
the true level in the simulation setup: The number of true source
dipoles is 20, and so the total number of active source components
is 60 after considering orientations. Thus, the true sparsity
level is $1-60/p= 1-60/15360 \approx0.996$.\vspace*{-3pt}

%

%
\begin{figure}[b]
\vspace*{-3pt}
\includegraphics{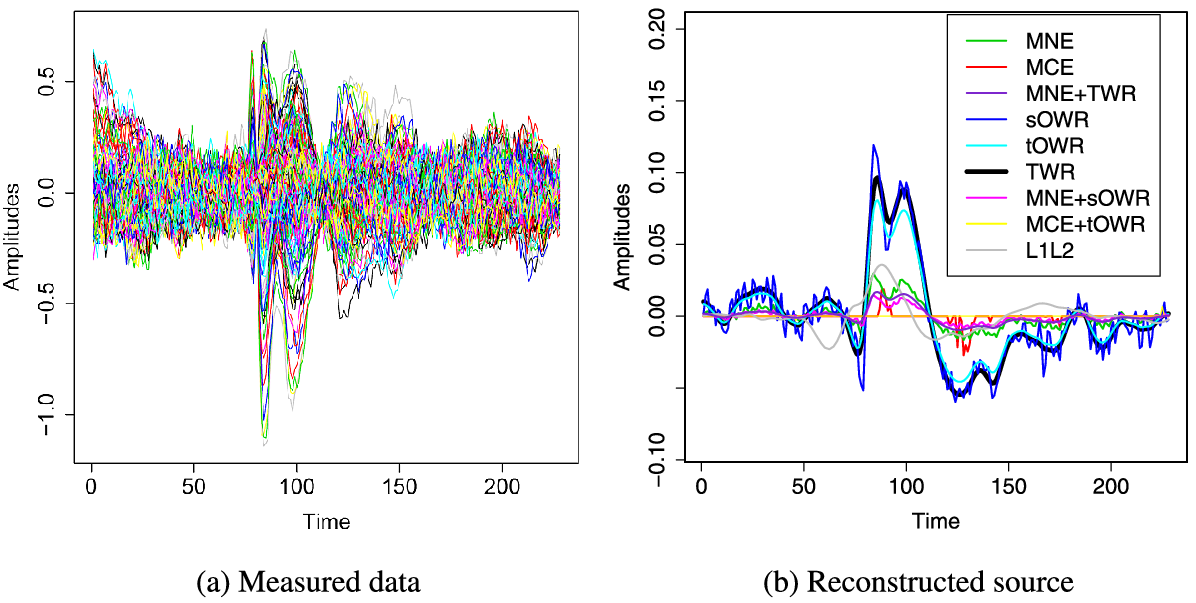}

\caption{MEG data. \textup{(a)} MEG recordings from 247 valid channels;
\textup{(b)} Reconstructed time courses from an arbitrary source
location in the somatosensory area by different methods.}
\label{figMEG}
\end{figure}

\section{Real data example} \label{secmeg}
In this section we demonstrate the proposed meth\-od using a human MEG
data set obtained from the Center for Clinical Neurosciences at
the
University of Texas Health Science Center at Houston. The study subject
is a
44-year-old female patient with grade three left frontal astrocytoma\vadjust{\goodbreak}
who underwent the MEG test as part of the presurgical evaluation. The
patient underwent a somatosensory task which is designed to
noninvasively identify the somatosensory areas of the patient. We
choose this study because of the clinical usefulness of the
somatosensory task in presurgical mapping.

Data collection was done with a whole-head neuromagnetometer
containing 248 first-order axial gradiometers. During the MEG
somatosensory session, 558 repeated stimulations were delivered to
the patient's right lower lip through a pneumatically driven soft plastic
diaphragm. Each stimulation lasted 40 ms with 450 ms epoch duration
(including a prestimulus baseline of 100 ms) and an interstimulus
interval randomized between 0.5~s and 0.6~s. We removed the offset and
averaged the 558 epochs to obtain the final event-related magnetic
field response. Then a bad channel was removed. The MEG
device recorded 228 time points in each epoch. The measurement
matrix $\bY$ is $247 \times228$, where $n=247$ is the number of valid MEG
channels and $s=228$ is the number of recorded data points per epoch. The
$n \times p$ forward operator $\bX$ was obtained using the MNE software
with $p=15\mbox{,}372$.

The measured MEG recordings from the 247 valid channels are plotted in
Figure~\ref{figMEG}(a). Among the 228 time points, there are two
peaks at time points 85 and 99, corresponding to the activation of
the primary somatosensory area contralateral to the stimuli,
as expected by clinical experiences and brain anatomic theories.

Nine methods, MNE, MCE, TWR, tOWR, sOWR, MNE+TWR,\break MNE+sOWR, MCE+tOWR
and $L_1L_2$, were applied to solve the MEG inverse problem. Figure
\ref{figMEG}(b) shows the reconstructed time courses for an
arbitrary source location by different methods. As we can see,
TWR, sOWR and tOWR, are satisfactory in terms of estimating
the shape of the source time course and capturing the peak features
at time points 85 and 99. But sOWR produces a noisy time course. MNE
and MNE+sOWR overshrink the magnitudes in addition to producing a noisy
time course. MNE+TWR recovers the shape of the time course but
underestimates the amplitude. The $L_1L_2$ method does not distinguish
the two peaks. MCE only identifies the first peak but misses the second
one. MCE+tOWR does not capture any activity because it smoothes the
spikes caused by MCE and hence is the least satisfactory method.

\begin{figure}

\includegraphics{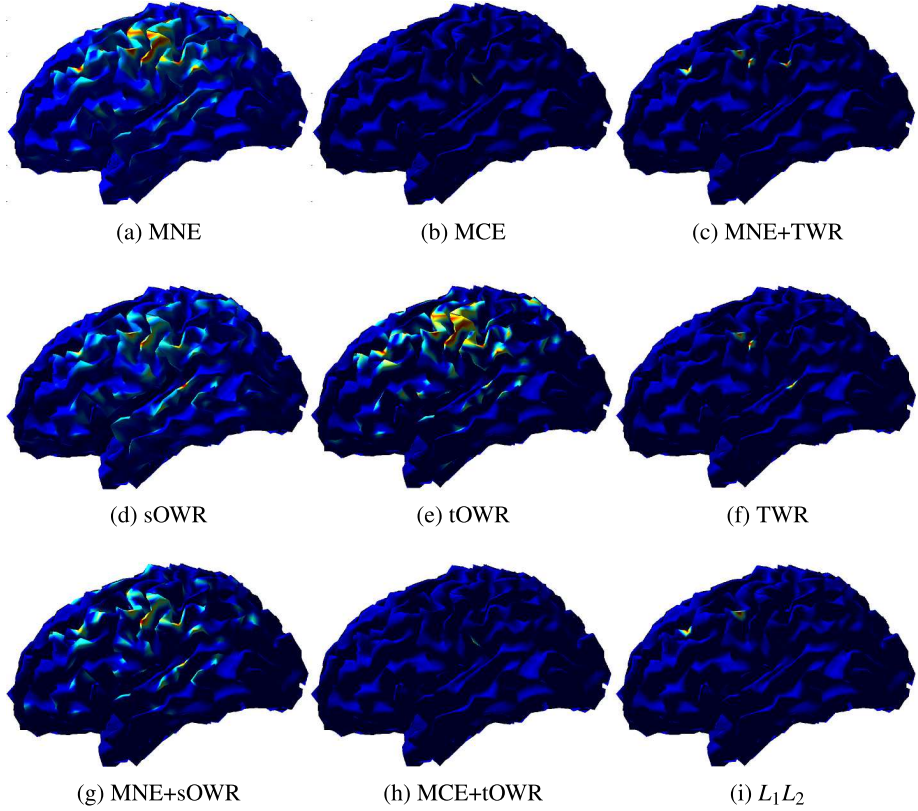}

\caption{Side views of the brain mapping at time point 85 by
different methods. TWR provides a focal and accurate detection; MNE+TWR
and $L_1L_2$ identify some activity in the frontal lobe in addition to
the somatosensory area. Solutions from MNE, sOWR, tOWR and MNE+sOWR are
too diffuse to be satisfactory. Both MCE and MCE+tOWR miss the activity
in the somatosensory area.}
\label{figmegmaps}
\end{figure}

Figure~\ref{figmegmaps} shows the side views of the brain mapping at
time point 85 by different methods. As we can see, the somatosensory
area was correctly identified by TWR, which matches the clinical
expectation. As with the synthetic example, tOWR and MNE produce
diffuse solutions, leading to false positives around the somatosensory
area. sOWR produces a scattering solution and so does MNE+sOWR. MNE+TWR
and $L_1L_2$ also identify some activity in the frontal lobe. Solutions
from MCE and MCE+tOWR are too focal and do not cover the somatosensory area.

%

\begin{figure}

\includegraphics{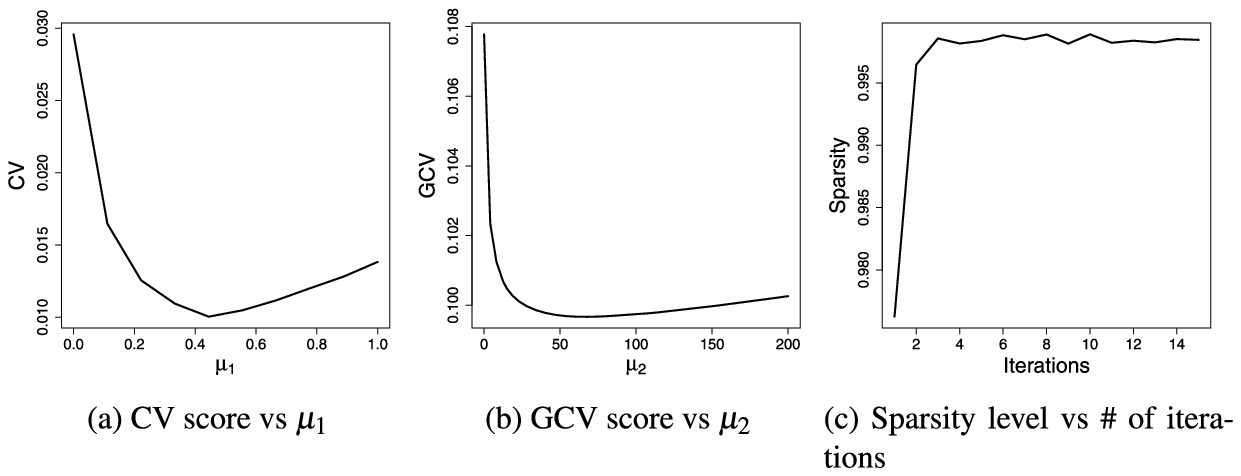}

\caption{Selection of $\mu_1$ and $\mu_2$ and the sparsity level
as a function of the number of iterations. The
optimal $\mu_1$ and $\mu_2$ are around 0.44 and 59.5,
respectively. The sparsity measure levels off at around 0.999.}
\label{figtunMEG}
\end{figure}

Figure~\ref{figtunMEG}(a) shows the CV error as a function of $\mu
_1$. The CV error was minimized when the sparsity parameter, $\mu_1$,
is about 0.44. Figure~\ref{figtunMEG}(b) displays the GCV error as a
function of $\mu_2$. It shows that the optimal $\mu_2$ is about 59.5.
The sparsity level as a function of the number of iterations is shown
in Figure~\ref{figtunMEG}(c). As we can see, the sparsity level
increases at first and then levels off rapidly, indicating the
algorithm converges fast. The optimal sparsity level was about 0.999.

%

\section{Discussion} \label{seccon}

TWR solves the MEG inverse problem by using two-way penalties that
promote both the temporal smoothness and the spatial focality of
the solution. We developed a computational efficient two-stage
procedure for implementing TWR. We also considered a one-stage
approach that tries to recover the source signal matrix $\bB=\bA\bG^T$
by solving
%
\begin{equation}\label{eqpen-1stage}
\min_{\bA,\bG}\{\|\bY-\bX\bA\bG^T\|_F^2+\mu_1|\bA|
+\mu_2\tr(\bG^T\bOmega\bG)\}.
\end{equation}
The optimal matrices $\bA$ and $\bG$ can be obtained by alternating
optimization.
When fixing $\bA$ as $\hat{\bA}$, the optimal $\bG$ can be obtained
as in Algorithm~\ref{algTWR}, as described in Section~\ref{secalg}. When
fixing $\bG$ as $\hat{\bG}$, the problem (\ref{eqpen-1stage})\break becomes
%
\begin{eqnarray} \label{eqA1}
&&\min_{\bA}\{\|\bY-\bX\bA\hat{\bG}^T\|_F^2+\mu_1|\bA|\} \nonumber\\
&& \qquad =\min_{\bA}\{\tr[\hat\bG(\bY\hat\bG-\bX\bA)^T(\bY\hat\bG
-\bX\bA)\hat\bG^T]+\mu_1|\bA|\} \\
&& \qquad =\min_{\bA}\{\|\bY\hat\bG-\bX\bA\|_F^2+\mu_1|\bA|\},\nonumber
\end{eqnarray}
which is equivalent to $s$ different problems, one for each column of
$\bA$, namely,
\[
\min_{{\mathbf{a}}_j}\{\|\bY\hat{\mathbf{g}}_j-\bX{\mathbf{a}}_j\|^2+\mu_1|{\mathbf{a}}_j|\},
\qquad j= 1,\ldots, s,
\]
where $\hat{\mathbf{g}}_j$ is the $j$th column of the matrix $\hat{\bG}$.
Each of these problems is a~standard
LASSO regression problem [\citet{T96}] with over 10,000 variables.
Although efficient computational algorithms exist for the LASSO
regression, the fact that the LASSO problem needs to be solved a few
hundred times during each iteration of updating $\bA$ makes this
approach computationally unattractive. Developing
a scalable algorithm for the one-stage approach is an important
issue for its practical application and remains an interesting
research topic.

\section*{Acknowledgments}
The authors thank the Editor, the Associate Editor and two referees for
their comments,
which helped improve the scope and presentation of the manuscript.
The authors thank the MEG Lab at the University of Texas Health Science
Center Houston for providing the data.
In particular, thanks are due to Professors Andrew Papanicolaou and
Eduardo Castillo for their suggestions and comments.


%

\printaddresses

\end{document}